# Mapping three decades of intellectual change in academia

Daniel Ramage[1,2], Christopher D. Manning[1], Daniel A. McFarland[3]

[1] Department of Computer Science, Stanford University
[2] Google Inc.
[3] Graduate School of Education, Stanford University

**Abstract:**
Research on the development of science has focused on the creation of multidisciplinary teams. However, while this coming together of people is symmetrical, the ideas, methods, and vocabulary of science have a directional flow. We present a statistical model of the text of dissertation abstracts from 1980 to 2010, revealing for the first time the large-scale flow of language across fields. Results of the analysis include identifying methodological fields that export broadly, emerging topical fields that borrow heavily and expand, and old topical fields that grow insular and retract. Particular findings show a growing split between molecular and ecological forms of biology and a sea change in the humanities and social sciences driven by the rise of gender and ethnic studies.

Individual researchers are largely aware of the academic precedents that influence their work.  But at the scale of whole fields, the aggregate impact of one field on another, like computer science on genetics, has not been measured. This work develops a direct, quantitative measure of language incorporation across disciplines by the application of statistical text models to PhD dissertation abstracts.  Our model takes as input the text of an abstract, along with the list of academic areas under which each dissertation was filed, and generates probabilistic assignments that map the words in each dissertation to the closest related areas in academia, taking into account context, word sense, and missing area designations.  The result is a direct measure of *language incorporation*: which words in each dissertation can be attributed to which source areas in academia? To our knowledge, this is the first such overall mapping of inter-disciplinary language use in academia.[1]

Fig. 1 shows the resulting aggregate pattern of language incorporation across all pairs of academic fields since 1980.  Around the ring are the broad areas of Engineering, Physical and Mathematical Sciences, Biological Sciences, Earth and Agricultural Sciences, Social Sciences, and the Humanities. Subjects within each area are shown as bars. Each link is a measure of the extent to which language is incorporated (in either direction) and is colored by the field that sends more. This figure provides a stark visual representation of the gulf between the sciences and engineering, on the right, and the humanities and social sciences, on the left. Very few dissertations incorporate much language from across the divide.

---

[1] This paper was written in 2012, at the end of Daniel Ramage's Stanford Ph.D. His full dissertation is available at http://purl.stanford.edu/kw138fy5342.

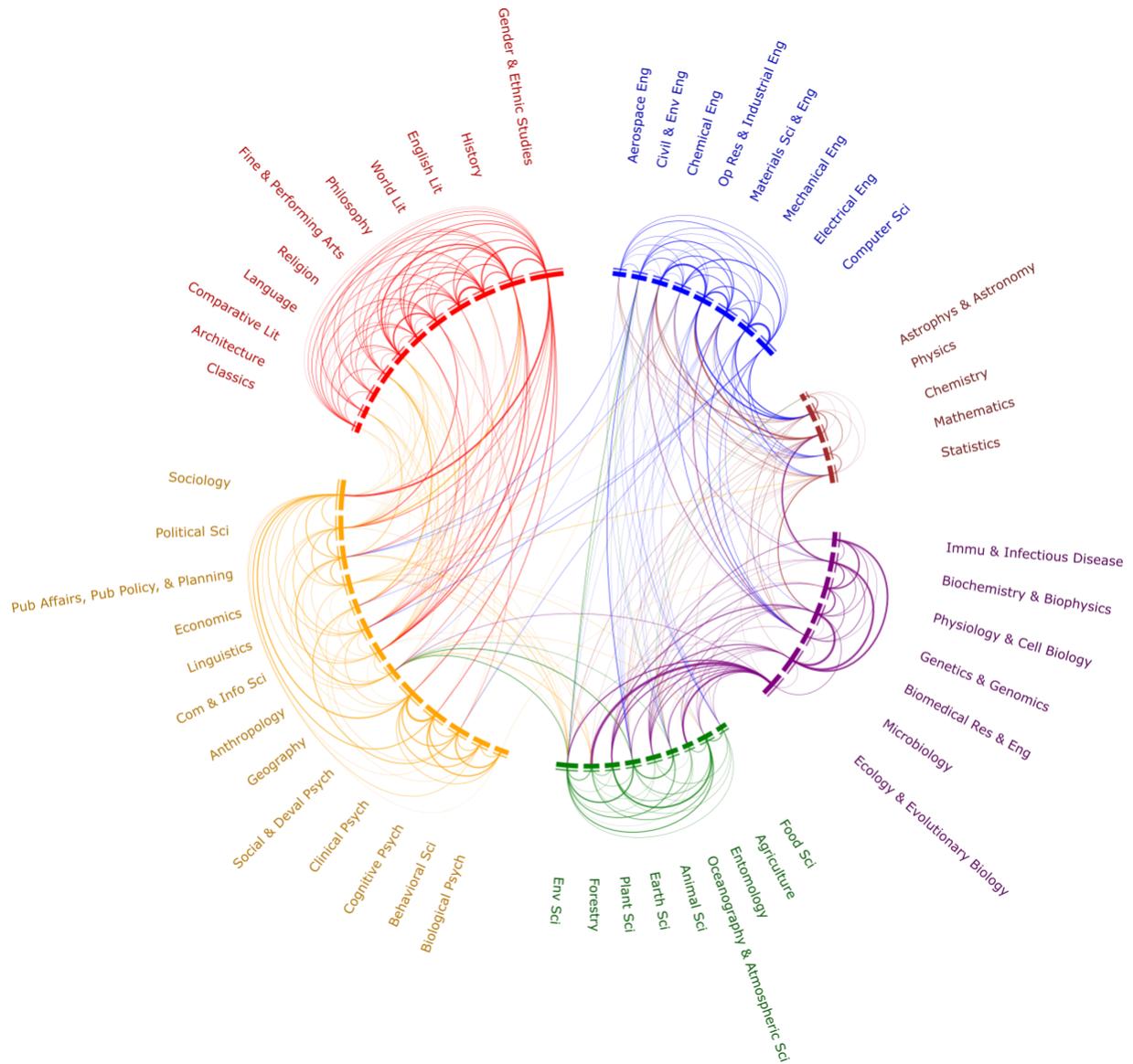

**Fig. 1: Language incorporation across all fields in academia. Every area is shown on the outer ring, grouped by broad area. Clockwise from top: Engineering, Physical & Mathematical Sciences, Biological Sciences, Earth & Agricultural Sciences, Social Sciences, Humanities. Arcs are drawn between areas with thickness in proportion to the total amount of language borrowed between the fields (in either direction) and with color determined by the area that sends more language. Arcs between broad areas are shown inside the circle, while arcs within a broad area are shown outside the circle, in order to emphasize broad multidisciplinary influences. Note the extent of division between the STEM fields (right) and non-stem fields (left).**

The model can discover the formation of new interdisciplinary areas. Fig. 2A shows the uptake of terms from Computer Science in dissertations from Genetics and Genomics and vice versa as a percentage of each field. While the growth in these areas' usage of each other tracks closely, Computer Science's usage of Genetics leads slightly in the early 2000's before Genetics' use of Computer Science reaches a higher peak in the late 2000s. The model tells us that interdisciplinarity is directional: the amount of genetics language incorporated by computer science is not equal to the amount of computer science language incorporated by genetics. Fig. 2B shows a starker example of asymmetric influence: ecology and evolutionary biology has had a larger impact on environmental sciences than the other way around. Indeed, asymmetries abound – some areas consistently incorporate more language from other areas than vice versa. Some areas act as language organizers for a broad area, such as Sociology for the social sciences or Ecology and Evolutionary Biology for the Earth and Agricultural Sciences.[2]

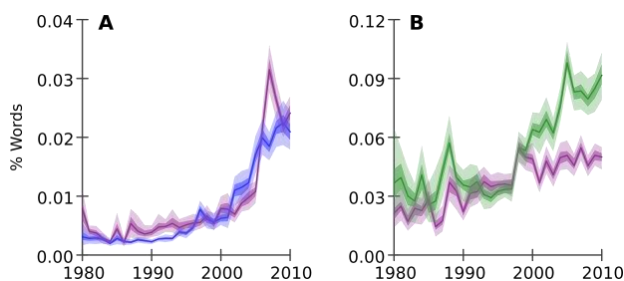

Fig. 2: Concurrent (A) and asymmetric (B) language incorporation in two pairs of fields. (A) shows the concurrent rise of computational biology (blue) and bio-computation (purple). The percentage of words in the computer science incorporated from genetics and genomics is shown in blue, while the percentage of words in genetics and genomics incorporated from computer science is shown in purple. Error bars are derived from bootstrap resampled estimates of the statistic (5% and quartile). (B) shows the asymmetrical incorporation of ecology and evolutionary biology into environmental sciences (green), versus the other way around (purple).

Prior academic study of scholarship has primarily used traditional methods from the social sciences including literature reviews, expert interviews, and surveys (*1, 2*). Most studies examine single fields (*3*) or compare several (*4*). In aggregate, these case studies convey stories about the development and division of broader areas, such as the divide between STEM fields (science, technology, engineering, and math) and the rest of academia (*5*); the growth of reductionism in general and microbiology in particular (*6*); and the growth in climate science and gender and ethnic studies (*7*). These observations are largely confirmed in the patterns of language incorporation we document. However, detailed studies employing traditional methods do not scale to similarly complete studies of academia writ large.

The few larger-scale studies of scholarship are based on link analysis rather than language use (*8*). These methods analyze networks of formal variables describing links such as citations or co-authorship (*9-11*) or interdisciplinary mixing in team science (*12*). However, network-based studies are limited to formal linkages, missing the hidden structure of academia that arises in informal conversations and often-uncited distant readings of others' work. As a result, they lack a representation of how concepts may be borrowed across disciplines. Furthermore, such approaches are limited by the availability of high quality, accurately disambiguated metadata that crosses field boundaries. Findings are therefore biased toward journal-heavy fields like biomedicine and fields with high-output short papers and short-term collaborations, under-representing book-heavy fields like the humanities. What is needed is an approach with uniform coverage of academia that does not overly rely on network link metadata.

---

[2] Further detail and information on materials and methods are available in the appendix.

**Method.**  We choose PhD dissertation abstracts for our study of interdisciplinarity because dissertations have greater coverage of research fields than do journal citation indices: most academics write one thesis that embodies several years' effort, and which generally extends the knowledge within a scholarly field.  We examine a subset of the UMI database maintained by ProQuest (*13*), analyzing 1.05 million dissertation abstracts filed between 1980 and 2010 from 157 schools classified as research-intensive by the Carnegie Foundation (*14*).  Our study dataset is thirty-one years: long enough for longitudinal analysis while ensuring high coverage of all schools and areas.  In addition to the title and abstract, dissertations in ProQuest are annotated by the authors with one of 268 common subject codes manually curated by ProQuest staff.  The subject codes in ProQuest are extensive, but much more fine-grained and with less clear organizational validity than the well-established basic disciplines reflected in common field designations like those in the National Research Council's 2010 report (*15*). Consequently, we grouped subject codes into 69 areas based on the NRC classification, which we, in turn, group into the six broad area designations introduced in Fig. 1.[2]

We associate the words in each dissertation abstract with their most similar area by using Partially Labeled Dirichlet Allocation (*16*), a Bayesian statistical text mining approach based on earlier statistical topic models (*17, 18*). We first learn a model for the kinds of language used in each field by exploiting the associations between words and areas.  Next, we re-examine each dissertation with respect to these area-language models, inferring the likelihood that each word could have been incorporated from some other field. The model discovers incomplete and variant usage of each label automatically, and is used to learn the vocabularies best associated with each area label across all documents. Details appear in the appendix.

Our method discovers well-known histories of interdisciplinarity among intellectual disciplines, such as the recent rise in computational biology (*6*).  In addition, we document new findings based on patterns of influence across many pairs of areas.  First, we find that fields play distinct roles in language production: some areas are net sources of language in other fields (methodological fields), while other fields mostly borrow language or build their own (topical fields).  Then, we describe two major patterns in multi-discipline dynamics over the past three decades, illustrated by the split in the biological sciences and the rise of gender and ethnic studies.

**Disciplinary Roles in Language Production.**  The flow of concepts across disciplinary boundaries is rarely balanced: fields vary in both the frequency and extent to which they export their language or incorporate the language of others.  Some fields, such as philosophy, provide conceptual guidance to others by exporting modes of inquiry.  Other fields, like those in the agricultural sciences, tend to apply and extend general principles in a topical domain.  One way to formally measure such aggregate differences in field-level language use is the *net source score S* for an area *a*: the number of times area *a* exports more language to other fields less the number of times *a* incorporates more language from other fields. Fig. 3A shows the net source scores in each broad area over time, with higher scores demonstrating higher relative influence.  The shape of these point clouds demonstrates that some areas in Engineering[3] and the Social Sciences have gained in influence, whereas many in the Physical & Mathematical Sciences, Earth & Agricultural Sciences, and some of the Humanities have lost influence. Selected areas are shown in Fig. 3B as trajectories of area size versus net source score over time.

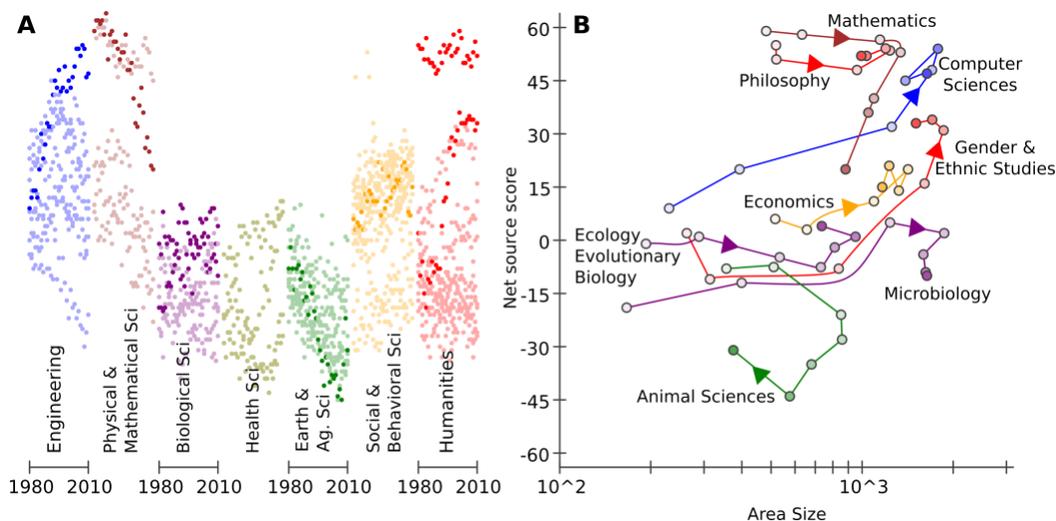

Fig. 3: Net source score (y, shared axis) for academic areas. In (A), each area's net source score is plotted over time, grouped by broad area. The highlighted areas are in detail in (B) where each area's size (x) is plotted versus its net source score over time (line series). In (B), from top to bottom, the brown line is Mathematics; the red lines from the Humanities are Philosophy and Gender and Ethnic Studies; the blue line is Computer Science, the Purple lines are Ecology and Evolutionary biology and Microbiology, and the Green line is Animal Sciences. Each line represents a time series from 1980 to 2010 by 5-year increments, progressing from the lightest to darkest dot. The results for Statistics rest atop Mathematics and Philosophy, so it was excluded for graph legibility.

The highest exporting areas tend to be methodological or concern abstract reasoning: e.g., mathematics, philosophy, computer science, and statistics. Mathematics and philosophy have elsewhere been described as "root disciplines" (*19, 20*) because they represent a fundamental form of knowledge with direct applicability in other fields. We find quantitative support for this observation: mathematics ($S = 32$) and philosophy ($S = 54$) are strong net sources. But as contemporary science and engineering become ever more data driven, the fields that enable the collection, processing, and interpretation of large datasets are coming to inhabit a similarly central role. Computer science ($S = 55$) and statistics ($S = 54$) anchor this new category of the data-driven root discipline.

Other areas do not act as net sources of language. These areas tend to be topically focused, including humanistic disciplines like classics ($S = -23$) and languages ($S = -19$) or applied disciplines like earth and agricultural sciences (average $S = -25.9$). Even areas that have grown in size and importance, such as the biological sciences, do not consistently act as sources of language with respect to other fields. While tripling in size over the past decades – from roughly 2,000 dissertations per year in the 1980s to more than 6,500 per year in the 2000s – the biological sciences still borrow more language from other areas, averaging a negative net source score of $S = -13$.

---

[3] Not all areas of Engineering gain influence: the gains are driven largely by increases in Computer Science, Electrical Engineering, and Operations Research. The notable downward trajectory in Fig. 3A is Chemical Engineering.

The division between exportable methods and topical applications suggests that some fields have an inherent advantage in the competition between disciplines in the "hierarchy of attention" (*7, 20, 21*). Our language-driven method indicates which kinds of knowledge are gaining influence and suggests new metrics for effectiveness of funding allocation. For example, more funding toward molecular biology might explain growth in that field and much practical advancement in the area, but it has not resulted in sustained methodological impact outside of the biological sciences. Indeed, biology's expansion has been enabled, in part, by incorporating methods from elsewhere: biophysics, biostatistics, bio-computation, bioengineering, bioinformatics, etc. A similar trend may eventually play out in neuroscience.

Overall the study has intriguing results at the level of individual fields (environmental science moves in this period from being a borrower to an exporter) and of broader disciplines (interdisciplinarity does increase in all areas, and most strongly between the 1980s and the 1990s). Below, we focus on two cases.

**The Rise of Molecules and Machines.** In the 1980s, the biological sciences were dominated by two primary modes of inquiry: (i) integrative approaches to biological systems, from individual animals all the way up to ecosystems, and (ii) reductionist approaches that sought to understand biology from its base components, first through microbiology and later through more specific fields such as genetics and genomics as well as cell biology (*6*).

Fig. 4 shows borrowing among all the biological sciences, health sciences, and earth and agricultural sciences in the 1980s and 2000s. Ecology and evolutionary biology – once an integral part of the biological sciences both by mass and by the extent its language was incorporated elsewhere – shrinks greatly in influence within the rest of the biological sciences. However, over the same period of time, it increasingly influences the earth and agricultural sciences. Simultaneously, reductionist biological fields grow within the biological sciences. Fig. 6 shows the amount of language exported by each of several areas to other areas within selected broad areas. For instance, we see that microbiology grows quickly through the 1980s before losing relative impact in biology to related reductionist approaches in the 1990s, including genetics and genomics as well as physiology and cell biology. Molecular biology acts as an integrator of the rest of the biological sciences and a growing source of language for health sciences, as well. In a sense, this split reflects a fundamental divide latent in the unit of analysis and types of explanations and epistemologies practiced within the two halves of the biological sciences. The majority of the field of biology is now linked to reductionist approaches and applications in medicine, while ecology and evolutionary biology has nearly split off into the emerging domain of environmental studies with applications and influence in the earth and agricultural sciences.[4]

---

[4] Indeed, the split in the field is so deep that some universities have already discussed dividing their departments into two if they have not done so already – molecular biology and evolutionary biology (*23*).

The ability to quantify and document these trends has implications for the way we structure university initiatives. For instance, many consider the rise of environmental studies as the de novo birth of an independent discipline (*22*), but it heavily incorporates language from ecological and evolutionary biology.  In contrast, the rise in reductionist biology is driven by technological miniaturization, data generation, data collection, and associated engineering methodologies.  Indeed, the percentage of language in the biological sciences borrowed from engineering disciplines as a whole roughly doubles over the time period to about 7%.

**The Rise of Gender and Ethnic Studies.**  The origins of gender and ethnic studies predate our dataset's first year to the 1960s.  At that time, the humanities and social sciences faced an intellectual crisis of legitimacy in an era of broad cultural shifts that demanded greater representation of traditionally under-represented voices and views (*5, 7*).  Over the next decades, these forces built a foundation for phenomenal growth in influence of gender and ethnic studies.  Fig. 5 documents this trend, showing language borrowing among the social sciences and humanities over our time span.  Starting in the late 1980s, the influence of gender and ethnic studies grows rapidly (with institutional support (*5*)), as does the size of the field.  By the 1990s, gender and ethnic studies has become an organizing force throughout the humanities and social sciences, with its influence only beginning to plateau in the early 2000s.

Contemporary with the rise in gender and ethnic studies, philosophy declines in size but not in relative influence, remaining a strong net source of language (Fig. 3), reflecting the continued strength of traditional modes of inquiry in the humanities and social sciences.  The growth of gender and ethnic studies acts as a complement to the traditional methods, demonstrating the increasing relevance of critique (sometimes referred to as an "antidisciplinary" stance (*5*)) and awareness of the role played by identity in humanistic and social scientific inquiry. Perhaps unsurprisingly, we find no evidence of a concurrent rise in language usage of gender and ethnic studies in either engineering or the physical and mathematical sciences.

Long ago, Thomas Kuhn argued that scientific paradigms were defined by their language (*1*). We now have a quantifiable means of analyzing language dynamics to see how words define fields and the way they change. We present one such approach and use it to describe inter-field dynamics in specific fields and broader areas of study.  However, these methods only illustrate what has happened; they leave open causal and counterfactual questions such as what would have happened if NIH had funded Systems Biology more strongly, or whether the decline of cognitive science work linking Computer Science and Psychology was or was not for funding reasons.

**Acknowledgements:**
This work was supported by the National Science Foundation under Grant No. 0835614.

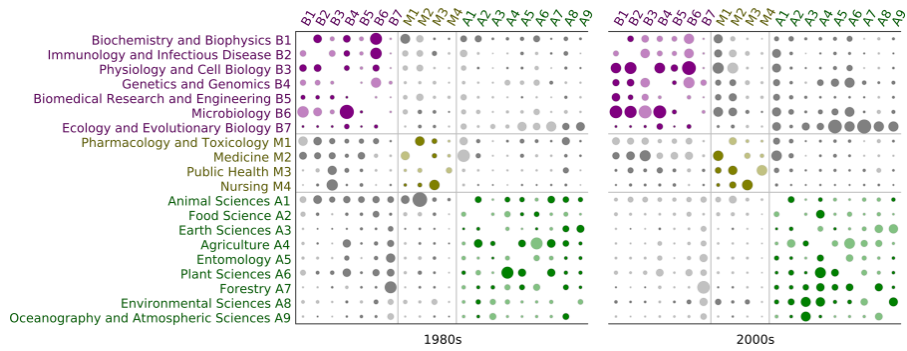

Fig. 4: Interdisciplinary language incorporation among the Biological Sciences (purple), Health Sciences (gold), and Earth and Agricultural Sciences (green) at two time points (1980s and 2000s). The value of cell $i,j$ represents the fraction of words in column $j$ that were incorporated from row $i$. Note the tremendous increase in language in the Earth & Agricultural sciences incorporated from Ecology and Evolutionary Biology. Note also the increased influence of the Biological Sciences (as compared to the Animal Sciences) on Health Sciences.

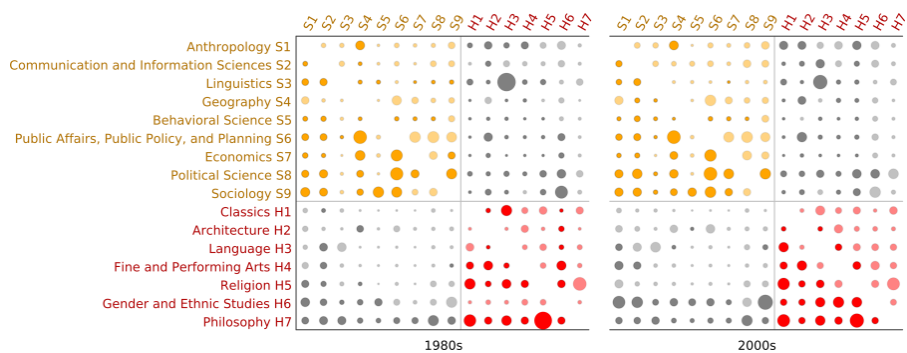

Fig. 5: Interdisciplinary language incorporation among the Social Sciences (orange) and Humanities (red) following the same conventions as Fig. 4. The major driver of change is the increased influence of Gender and Ethnic Studies across both the Social Sciences and Humanities.

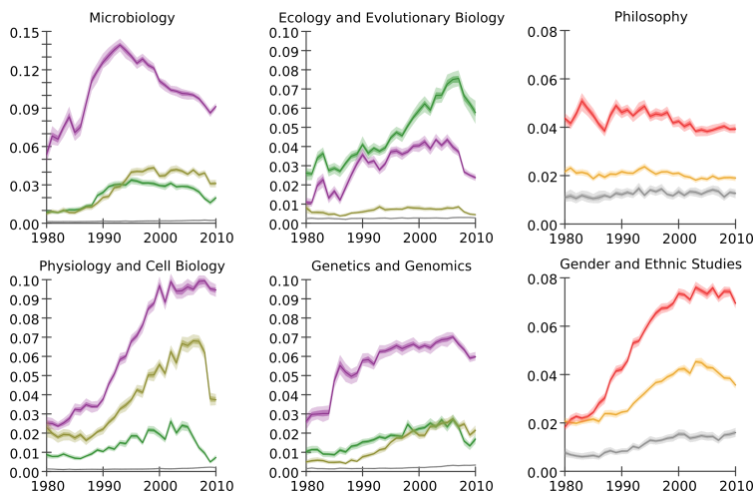

Fig. 6: Language incorporated from each area (graph title) over time from other areas of the Biological Sciences (purple), Health Sciences (gold), Earth & Agricultural Sciences (green), Humanities (red), and Social Sciences (orange). These graphs represent the total percentage of words in the given broad area incorporated from the named area. Usage in all other broad areas not captured by the lines in a given graph is shown in gray.

# Appendix: Materials and Methods

## Data

Shortly after completion, most PhD dissertations in the United States are filed in the UMI database maintained by ProQuest. ProQuest is designated by the Library of Congress as the collection agency for published PhD dissertations in the United States. The subset analyzed contains 1.05 million dissertation abstracts filed between 1980 and 2010 from 157 schools. We selected schools that have been classified as research-intensive in one of three surveys of higher education conducted by the Carnegie Foundation since 1994 (*9*). We examined only data from 1980 because electronic abstract records became much sparser before then.

Each dissertation contains a title, abstract, author, advisor, date, subject codes, and keywords. The abstracts contain an average of 179 words after removing common stop-words (such as *about* and *the*), removing rare terms occurring in less than 5 documents, and collapsing term variations with a Porter stemmer (*24*).[5] 268 commonly used subject codes in our dataset are taken as labels. The subject codes reflect domain expertise and the knowledge of both the filer and the taxonomists at ProQuest, and they correspond with relatively high-level field designations such as biochemistry, public administration, cultural anthropology, etc. These subject codes are proposed by filers at the time of dissertation submission and are manually curated by ProQuest. Some subject codes are introduced or disappear during the time span of the data, but most are stable designations over the 31 year period.[6]

Most dissertations (92%) have more than one subject, with the bulk having either two (58%) or three (24%) subject labels affixed to them. Unfortunately, the subject codes themselves are unevenly distributed: some areas like Physics have a rich taxonomy of codes (13 subject codes for 52,432 dissertations) whereas other areas like Computer Science contain a paucity of codes (only two subject codes for 41,605 dissertations).

The subject codes in ProQuest are extensive, but more fine-grained and in only partial alignment with the well-established basic disciplines reflected in common field designations like those in the National Research Council's 2010 report (*15*). Consequently, we grouped subject codes into 69 areas based on the NRC classification, which we, in turn, group into six broad area designations: Engineering, Physical & Mathematical Sciences, Biological Sciences, Earth and Agricultural Sciences, Social Sciences, Humanities. Four more broad areas primarily oriented toward professional training – Education, Business, Law, and Health & Medical Sciences – are not considered in the analysis. Even after grouping subject codes into areas, we find that the average dissertation contains 1.6 areas, with nearly half (46%) still participating in more than one area and many (17%) having three or more areas designated.

---

[5] In general, stemming is not necessarily useful in topic modeling because synonymous variants are often correctly placed into the same topics. The reason terms are stemmed here is simply computational convenience: we are memory limited, so reducing model size – by reducing vocabulary size – allows more document data to fit each compute node.

[6] ProQuest has made changes to the subject code hierarchy over the time span of the data, creating mappings between hierarchies. Another difference by year is in the number of subject codes that can be applied while filing. The approach we take here controls for such variations (to a large degree) by filling in missing labels with Partially Labeled Dirichlet Allocation inference.

Subject codes were grouped using a two-phase methodology: automatic clustering followed by manual curating.  In the clustering phase, we computed a vector of term weights for each commonly used subject code, resulting in 269 vectors weighted by inverse document frequency (*25*), as is standard practice in text processing.  Then, we used single-link hierarchical agglomerative clustering (*26*) to build a tree of subject codes that reflect the bottom up similarity of pairs of subject codes.  In the curating phase, we manually cut the inferred subject code tree at levels corresponding to broader area designations in ProQuest and NRC, slicing it into 69 areas reflected in the NRC report.  Subject codes were moved when necessary to reflect these common designations.

## Statistical Topic Modeling

Words in each dissertation abstract are associated with the most similar area label by using Partially Labeled Dirichlet Allocation (*16*). The procedure is a two-step process: learning and inference.  We first learn a model for the kinds of language used in each field by exploiting the associations between words and areas.  Next, we re-examine each dissertation with respect to these area-language models, inferring the likelihood that each word could have been incorporated from some other field.

Partially Labeled Dirichlet Allocation (PLDA) describes the relationships between words, documents, and labels. In the learning phase, we build models of the language in each field by examining the co-occurrence of field designations and words in their corresponding abstracts.  From only a single dissertation with two or more labels, we could not hope to discern which words belong to each label. But by looking at the distribution of words and labels across the entire collection we can learn that words such as "genome" and "sequence" are statistically more likely to occur together in Genetics & Genomics documents, whereas terms like "algorithm" and "complexity" are better attributed to Computer Science.  As a result, we can determine which words in a dissertation labeled both as Computer Science and Genetics & Genomics are better attributed to each label (and, more specifically, what kind of Computer Science or Genetics). Inference is a context-sensitive soft clustering approach: a word like *sequence* can belong to both Mathematics and Genomics, and an instance of it is probabilistically attributed to one or both fields depending on the other words in the abstract. PLDA builds on recent work in the topic modeling community (*17*) that has been widely extended as well as applied to the study of large scale phenomena in science (*18*, *27*).

The majority of existing topic models are *unsupervised* models of text that discover latent *topics* of words that tend to co-occur in a document collection, as epitomized by the popular Latent Dirichlet Allocation model (*17*).  While the topics learned by unsupervised models on our document collection do have qualitative appeal, they do not explicitly align with the human-interpretable area labels applied to the dissertations.  As a result, we have no sense of what the right number of topics is, nor can we trust that the resulting topics signify trends in usage of specific area designations.  PLDA uses the machinery of LDA to model latent topics *within each area designation* – it learns which words statistically co-occur with each area designation, and the latent variations therein.  Table 1. Top terms in computer science overall (left) and by automatically discovered latent sub-area (right). shows the top terms in computer science overall and within each automatically discovered sub-area.

Table 1. Top terms in computer science overall (left) and by automatically discovered latent sub-area (right).

| systems design problem algorithm algorithms network approach techniques problems applications networks models method software methods set number | ← Top computer science terms | Top terms by latent sub-area → | databas queri web file access retriev storag user search document system relat process | design softwar user system applic environ interfac tool provid support implement | program languag gener logic specif formal semant code implement system |
|---|---|---|---|---|---|
| | | | algorithm problem graph comput optim solv solut number effici complex bound | learn cluster featur algorithm recognit classif mine train extract classifi set | network protocol rout servic commun distribut node propos applic mobil wireless |

Specifically, PLDA assumes that for each label $l$, there exists a set of $K_l$ topics. Each topic is itself a multinomial distribution $\beta_{lk}$ over the vocabulary $V$ describing some set of words that tend to co-occur within a document. While the labels are known, the topics are not – the model discovers variations in the usage of each label automatically. In this case, that means that it discovers sub-disciplines.

Formally, the likelihood that any given term $w_i$ at position $i$ in a document $d$ comes from any given label $l$'s topic $k$ is proportional both to how much that topic likes that term $\beta_{lk}$ times how much that document likes that label $\psi_{dl}$ and topic $\theta_{dlk}$. Formally, the probability of sampling a particular label $j \in L$ and topic $k \in K_j$ given some document's observed set of labels $\Lambda_d$ and Dirichlet hyper-parameters $\eta$ and $\alpha$ is given by:

$$P(l_{d,i} = j, z_{d,i} = k | l_{\neg d,i}, z_{\neg d,i}, w_{d,i} = t; \alpha, \eta) \propto$$

$$I[j \in \Lambda_d \wedge k \in 1..K_j] \cdot \left( \frac{n_{\cdot,j,k,t}^{(\neg d,i)} + \eta}{n_{\cdot,j,k,\cdot}^{(\neg d,i)} + V\eta} \right) \cdot \left( n_{d,j,k,\cdot}^{(\neg d,i)} + \alpha \right)$$

In the inference phase, we re-examine every dissertation without the restriction that its words be generated by one of the dissertation's labeled areas: i.e., $\Lambda_d$ is considered to be the full set of labels $L$. We allow the model to determine the optimal mix of areas that would result in the words we see in any given dissertation. The resulting distribution can be interpreted as the percentage of words in a given dissertation (or set of dissertations) that can be attributed to some other field.

### Net source scores

The net source score is a sum over all other areas $b$, adding one if $b$ incorporates significantly more language from $a$, or subtracting one if $a$ incorporates significantly more language from $b$. We compute statistical significance by comparing the distributions of our statistic (mean language use of some area) in 500 bootstrap samples of the dissertations in any target area at the 5% confidence interval. Note that not every area contributes to the sum: many pairs of areas incorporate each other's language at statistically indistinguishable rates.

## Validating results

Because our analysis is with respect to a given categorization scheme (the labels $L$), the specific way in which the labels are defined implies differences in the languages learned within each label. To validate our model choices, we use two different, reasonable categorization schemes (ProQuest subject codes and their aggregation into NRC areas as described in the **Error! Reference source not found.** section). We look to where they converge in their view of the document collection, suggesting an appropriate level of dimensional definition that best reflects latent dimensions of language organization while still corresponding with the expert information afforded by two respected classification schemes.

To derive real qualitative insight into the data, we need to be sure that the statistics the model computes reflect real patterns of language usage in the world and not just artifacts of a choice of parameters or classification. To select the number of topics per model, we fit many models to the data varying the number of latent subjects per area designation, looking for areas of consistency between the models' assessment of cross-disciplinary language incorporation. Fig 7 shows the agreement between models based directly on subject codes and models based on areas (aggregated subject codes) as the number of latent topics per label varies. Note the high absolute correlations overall – representing the relative stability of the models learned – as well as the tendency for models trained with a sufficient number of latent topics under either labeling scheme to agree. The agreement demonstrates that the patterns of language incorporation learned are stable with respect to variations in the number of topics per label (for sufficiently many labels) as well as the granularity of the classification scheme. Where not otherwise stated, we use the 12 topics per area model because of its high consistency with the other models and its comparatively small size. The model has a total of 829 topics versus a maximum of 4,289 topics for the 16-topics-per-subject model. Learning and inference in the PLDA model can be accomplished in parallel. As a result, a model with 12 topics per label and 69 area labels can be learned in about a day on a small compute cluster. The analyses we present here are consistent across all models with from 8 to 16 topics per label.

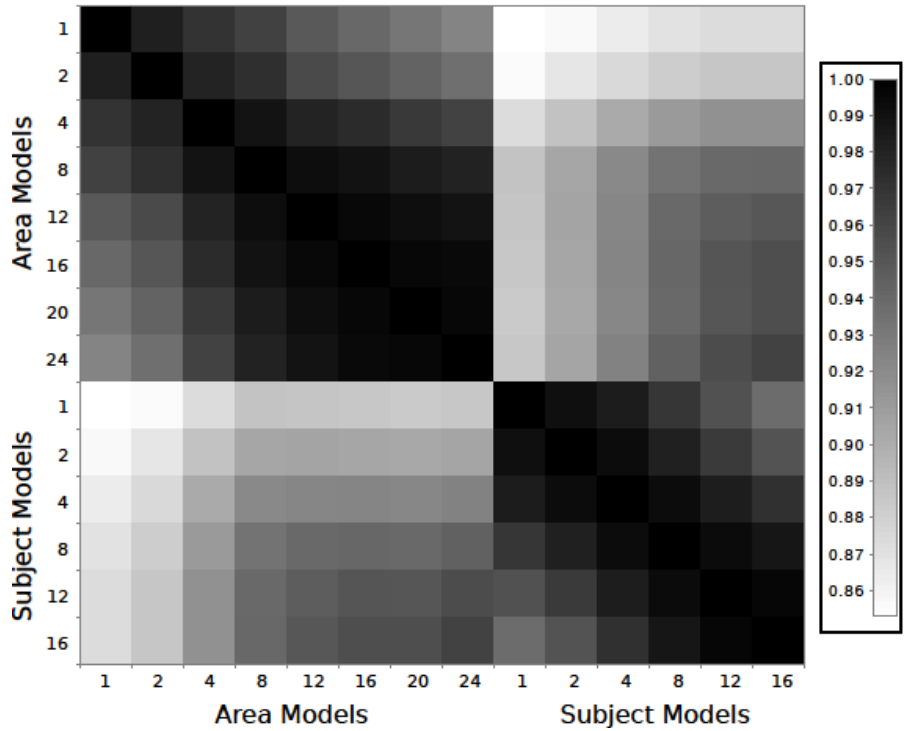

Fig 7. Intra-model consistency among PLDA models of academic fields. PLDA models were trained using one background topic, and one of 2, 4, 8, 12, or 16 topics per Subject (bottom, right) or per Area (top, left, also including 20 and 24 topics per area). These models are compared by first computing the expected percentage of words borrowed between all pairs of areas in all years. For the subject codes, the percentages of each subject are summed by area to create a comparable scale. Correlations among these inter-area borrowing percentages are computed for all pairs of models, generating the plot above.